\documentclass[twocolumn,pre,tightenlines,superscriptaddress,showpacs]{revtex4-2}
\usepackage{graphicx}
\usepackage{amsmath}
\usepackage{caption}
\usepackage{ragged2e}
\DeclareCaptionJustification{justified}{\justifying}
\usepackage{subcaption}
\usepackage{xcolor}
\usepackage[normalem]{ulem}

\begin{document}

\preprint{}

\title{Learning to flock in open space by avoiding collisions and staying together}


\author{M. Brambati}
\email[]{mbrambati1@uninsubria.it}
\affiliation{DISAT, Universit\'a degli Studi dell’Insubria, Como, Italy}
\affiliation{I.N.F.N. Sezione di Milano, Milano, Italy}

\author{A. Celani}
\email[]{celani@ictp.it}
\affiliation{The Abdus Salam International Center for Theoretical Physics (ICTP), Trieste, Italy}
\affiliation{Department of Oncology, Universit\`a degli Studi di Torino, Italy}

\author{M. Gherardi}
\email[] {marco.gherardi@unimi.it}
\affiliation{Universit\'a degli Studi di Milano, Milano, Italy}
\affiliation{I.N.F.N. Sezione di Milano, Milano, Italy}

\author{F. Ginelli}
\email[] {francesco.ginelli@uninsubria.it}
\affiliation{DISAT, Universit\'a degli Studi dell’Insubria, Como, Italy}
\affiliation{I.N.F.N. Sezione di Milano, Milano, Italy}


\date{\today}

\begin{abstract}
We investigate the emergence of cohesive flocking in open, boundless space using a multi-agent reinforcement learning framework. Agents integrate positional and orientational information from their closest topological neighbours and learn to balance alignment and attractive interactions by optimizing a local cost function that penalizes both excessive separation and close-range crowding. The resulting Vicsek-like dynamics is robust to algorithmic implementation details and yields cohesive collective motion with high polar order. The optimal policy is dominated by strong aligning interactions when agents are sufficiently close to their neighbours, and a flexible combination of alignment and attraction at larger separations.
\end{abstract}


\maketitle

\section{Introduction}
The synchronized flight of bird flocks, exemplified by starling murmurations, is perhaps the most striking example of collective behavior in natural systems, which fascinated scholars for quite a long time \cite{rackham1942pliny}. Evolutionary biologists, for instance, have long debated the advantages of living in groups \cite{parrish1999complexity}, which should offer increased protection from predation by diluting the individual risk and possibly confusing the attackers by the sheer size of the assembly. \\

Flocking behavior involves a high degree of order in the individual directions of motion \cite{cavagna2010scale}, and has been reproduced by minimal models of self-propelling particles (SPPs), such as Craig Reynolds Boids \cite{reynolds1987flocks} or the celebrated Vicsek model \cite{vicsek1995novel} that has long captivated the attention of statistical physicists and played a pivotal role in the birth of the active matter research field. The essential ingredient of these models is the tendency of individual particles to align their direction of motion with those of their local neighbours, which is enough to promote long range order in systems with finite density (even in two spatial dimensions, due to the non-equilibrium nature of self-propelled particles) such as in toy models with periodic boundary conditions. In open systems, constituted by a finite number of individuals in an open, infinite space, purely alignment interactions are however not enough to maintain group cohesion. Models with metric interaction rules, such as the original Vicsek model, will quickly disintegrate under the action of fluctuations\cite{ginelli2016physics}, while more realistic topological interactions will see an endless diffusive spreading of the group size \cite{ginelli2010relevance}. Therefore, it has long been recognized \cite{gregoire2004onset, chate2008modeling}, that some attractive interactions -- by which SPPs tend to orient their direction of motion towards their local neighbours position rather than their orientation -- are also needed to maintain group cohesion in open space. Typically, the relative weight of alignment and attractive interactions varies with the distance to neighbours, either sharply, as in behavioral zonal models \cite{couzin2002collective}, or continuously as in \cite{gautrais2012deciphering}. Alternative mechanisms capable of generating an effective attraction have also been proposed in the literature \cite{pearce2014role, giraldo2025active}.

Flocking, however, is not the only way for moving individuals to stay in a cohesive group, and also milling and swarming behaviors are observed in nature. If milling -- which involves SPPs collectively rotating in a single vortex \cite{couzin2002collective} -- still involves local aligning interactions and a form of global order (captured by a global normalised angular momentum), group cohesion can be also -- perhaps more easily -- achieved through local attractive interactions only, resulting in a globally disordered but compact swarming behavior. \\

It is therefore natural to wonder why certain gregarious species achieve group cohesion via flocking rather than swarming behavior, especially considering that flocking behavior is often displayed by animal groups that are not migrating or otherwise moving between different locations, such as starling murmurations. It has been argued that the flocking state -- the result of the spontaneous symmetry breaking of a continuous symmetry -- is characterized by long-ranged correlations \cite{cavagna2008new} and is thus highly susceptible and capable of a collective reaction to environmental perturbations such as predatory threats. However, quantitative studies of swarming midges show that they can display strong collective behaviour despite the absence of global order, with a large capability to collectively respond to perturbations which is comparable to that of highly ordered flock of birds. The biological function of and the biological imperatives leading to flocking as opposed to swarming collective behavior thus remains largely unknown. \\

In this work, we tackle this problem through the lens of Multi Agent Reinforcement Learning (MARL) \cite{barto2021reinforcement} considering both a centralized learning scenario, in which all particles share their training, and a concurrent one where each agent learns separately. This approach was explored in \cite{durve2020learning} where it was demonstrated that velocity alignment emerges as an effective strategy to favor cohesion. These findings were corroborated by a theoretical study of mean-field optimal stochastic control in \cite{borra2021optimal}. Here we extend the scope of these work by considering a richer model for the agents' behavior and their interactions.
Our SPPs move in a two dimensional open space and only perceive the positions and orientations of a local set of spatially balanced topological neighbours \cite{ballerini2008, ginelli2010relevance, camperi2012spatially}. They judge their current state in terms of the mean distance to these neighbours, and take actions to modify their instantaneous direction of motion, combining alignment with the neighbours orientation and attraction towards their position. For every state, they are trained to find the optimal combination of alignment and attraction according to a cost function based on the distance to local neighbours.  Regardless of the training being centralized or concurrent, we discover that cost functions that simply penalize large distances lead to a swarming behavior dominated by attraction, while the addition of a further penalty at small distances favor strong alignment at small to moderate distances, leading to a clear flocking and cohesive collective behavior similar to the one observed in starling flocks. \\

Our result, robust to the presence of a sufficiently small noise term in the orientation dynamics, demonstrates that the need to maintain a minimum distance from neighbours is essential for the onset of flocking behavior, suggesting that such collective behavior may emerge from two different imperatives: the urge to stay together but also the need to avoid potentially dangerous collisions between group members.

\section{Methods}
\subsection{Dynamics}
We consider $N$ agents moving off-lattice in a $2$-dimensional unbounded (that is, infinite) space. The dynamics is parallel and discrete and the agents update their position with the standard streaming rule
\begin{equation}
    \textbf{r}_i^{t+\Delta t} = \textbf{r}_i^t + v_0 \textbf{s}_i^{t+\Delta t} \Delta t,
    \label{r_t}
\end{equation}
where $\textbf{r}_i^t$ is the position of the i-th agent at time t, $\textbf{s}_i^t$ its unit heading direction and the speed of the agents is kept constant and equal to $v_0$.
In the following, without loss of generality, we put $\Delta t=1$. Agents only perceive their local environment represented by their topological (i.e. Voronoi) neighbours, the set ${V}_i$, a spatially balanced version of metric free interactions that grants optimal stability \cite{camperi2012spatially}. They integrate this information into the local average heading direction
\begin{equation}
    \textbf{V}_i^t = \Theta\left[\sum_{j\in \mathcal{V}_i}\textbf{s}_j^t\right]
    \label{Vi}
\end{equation}
and direction to the center of mass of the neighbours
\begin{equation}
    \textbf{R}_i^t = \Theta\left[\sum_{j\in \mathcal{V}_i}(\textbf{r}_j^t-\textbf{r}_i^t)\right]
    \label{Ri}\,,
\end{equation}
\begin{figure}[t!]
  \includegraphics[width=0.98\linewidth]{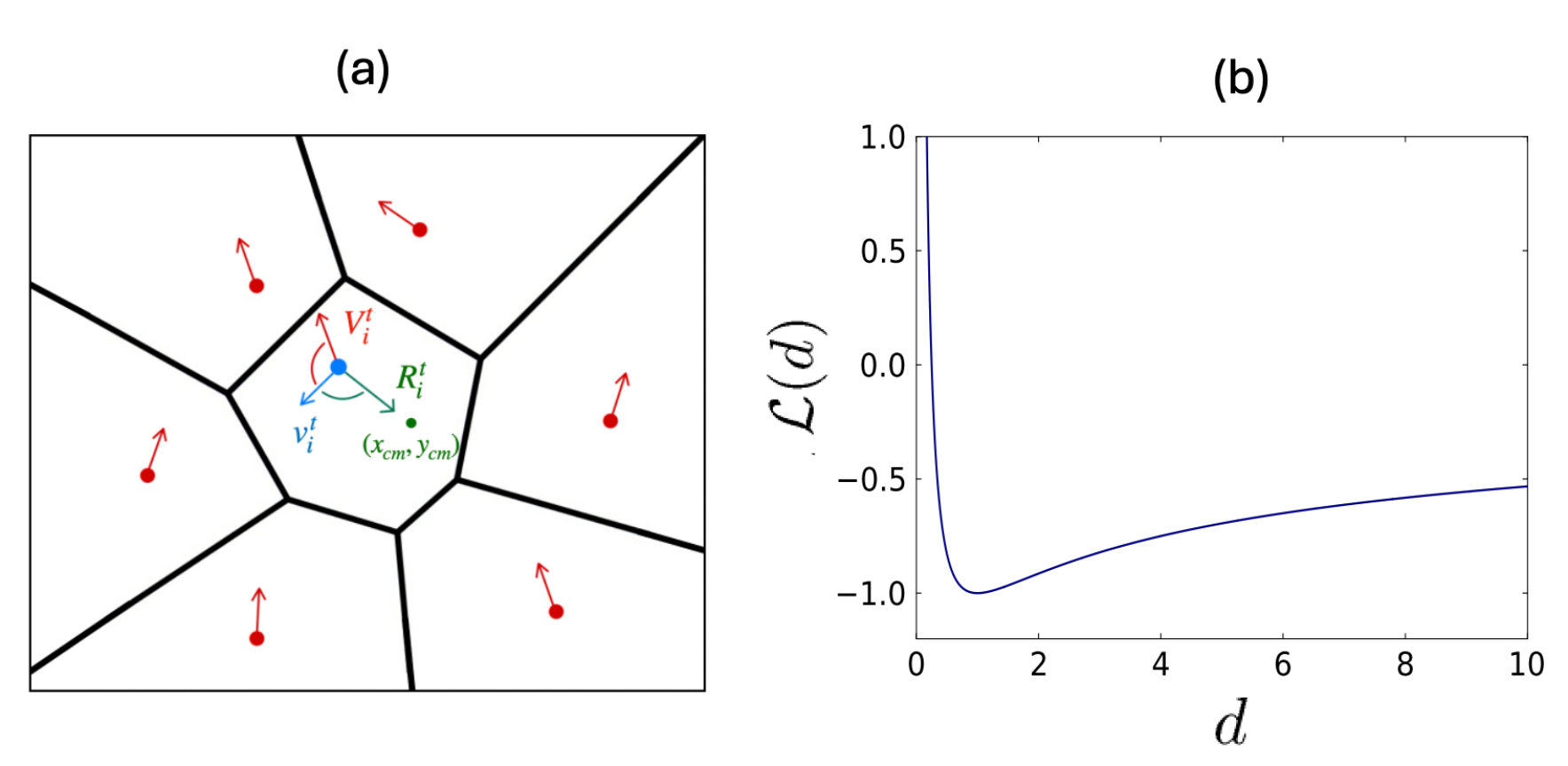}
\caption{(a) Cartoon describing the orientational dynamics of agents. At each step the $i$-th agent (blue arrow) chooses the new direction given the information it perceives from its topological neighbours (red arrows) and linearly weights the alignment to the neighbours' mean direction $V_i^t$ and to the direction $R_i^t$ to the local center of mass. Nearest neighbours are evaluated via Voronoi tessellation. (b) The cost function \eqref{L_x_eq} with parameters $a=1$ and $b=2$. 
} 
\label{graphic_summary}
\end{figure}
where the sum is over the set  $\mathcal{V}_i$ of Voronoi neighbours of the i-th agent.
These vectors are normalized to unity by the normalization operator $\Theta[{\bf v}]\equiv {\bf v}/|{\bf v}|$.
At each time step, agents change their direction of motion, aligning towards a combination of the two vectors above, weighted by a (learnable and state-dependent) relative coupling strength parameter $\beta \in [0,1]$, with 
\begin{equation}
    \textbf{s}_i^{t+1} = \Theta\left[(1-\beta)\textbf{V}_i^t + \beta \textbf{R}_{i}^t + \eta \xi_i^t\right],
    \label{s_t}
\end{equation}
where $\xi_i^t$ is a random unit vector with zero mean and delta correlations,
\begin{equation}
    \langle \xi_i^t \cdot \xi_j^{t'} \rangle = \delta_{ij} \delta_{t,t'}\,.
\end{equation}
Agents first update their heading via Eq. (\ref{s_t}) and then their position by Eq. (\ref{r_t}), a sketch of the dynamics is given in Fig.~\ref{graphic_summary}a. Note that for $\beta=0$ this rule reduces to the standard Vicsek model with topological interactions \cite{ginelli2010relevance}, while $\beta=1$ gives complete attraction towards the center of mass of local neighbours, that is
an attraction dominated dynamics. The relative coupling strength is thus the crucial parameter that will be optimized via the reinforcement learning algorithm described in the next subsection. Note finally that attractive forces are expressed as an (overdamped) torque acting on the particles heading as in Ref. \cite{gregoire2004onset}, so that particles heading always coincides with their direction of motion, a framework suitable for active agents that interact via social forces without physical collisions.
\subsection{MARL Framework}

Reinforcement Learning (RL) is a machine learning paradigm in which an agent learns a decision-making strategy through repeated interaction with an environment \cite{barto2021reinforcement}. At each time step, the agent observes a representation of its current state, which encodes relevant information about the environment, and selects an action based on this observation. The chosen action changes the agent state, after which the agent receives a scalar feedback signal—called a cost (or, equivalently, a reward)—that depends on the previous state representation, the action taken, and the resulting new state representation. This is realized through a cost/reward function ${\cal L}$ that encodes the agent’s overall objective and is used to reinforce or penalize the likelihood to select the same action when the same state representation is encountered again in the future.   
Eventually, repeating this cycle and through trial-and-error, an agent should learn an optimal {\it policy}, which maps state representations to actions, maximizing the cumulative reward (or, equivalently, minimizing the cost) over time. In multi-agent RL, different agents interact between themselves and with the environment. See Refs. \cite{monter2023dynamics, durve2020learning, wang2023modeling, loffler2023collective, cai2025reinforcement} for previous applications to active matter dynamics.

In this work we apply two different learning paradigms: centralized training (CT) and decentralized training (DT). In the former, agents act independently one from the other, but learn by shared experience -- as if they were \textit{hive-minded} -- while in the latter each agent learns its own policy independently, based solely on the sequence of states, actions and reward experienced by that specific agent.

The raw observation of each agent $i$ is given by the relative positions and headings of its Voronoi neighbours, while its {\it policy observation} (i.e. the information on which the policy is conditioned) is given by the mean distance of agent $i$ to its local neighbours,
\begin{equation}
    d_i = \frac{1}{m_i}\sum_{j\in \mathcal{V}_i}|\textbf{r}_j^t-\textbf{r}_i^t|,
    \label{eq:po}
\end{equation}
where $m_i$ is the cardinality of the set $\mathcal{V}_i$ of Voronoi neighbours.
Actions amount to the choice of a specific value for the relative coupling strength $\beta$ with which the agent performs one step of the dynamics (\ref{s_t})-(\ref{r_t}).

With a slight abuse of terminology, in the following we will use the term ``state" for the policy observation $d(i)$ (not to be confused with the {\it microscopic agent state} $({\bf r}_i, {\bf s}_i)$, and denote the chosen $\beta$ with the term ``action". This is a standard use in RL where one speaks generically of state/action pairs.
For the sake of simplicity,
we also discretize both the states and the actions 
in order to have a finite number of states $n_s$ and a finite number of possible actions $n_a$, avoiding the unnecessary computational complexity associated with continuous states and actions (we describe the details of the discretization in the Results section) . 

State-action pairs are thus a discrete set of pairs $(d,\beta)$
defining the entries of the so-called $Q$-matrix that in RL represent the estimate for the expected cost for each state-action pair and thus determines the optimal policy

\begin{equation}
\beta^*(d) = \operatorname*{arg\,min}_\beta Q(d,\beta)\,.
\label{optimal}
\end{equation}

We first discuss how the cost is encoded in the Q matrix and afterwards how actions are selected during learning. The goal we impose to our agent group is to maintain cohesion while, at the same time, avoiding collision risks between agents (that is relative distances getting too small). Therefore, in order to evaluate the performance of the state-action pair, we adopted a cost function in the form of a gentler version of a Lennard-Jones-like potential:
\begin{equation}
     \mathcal{L}(d) = \frac{a}{d} - \frac{b}{\sqrt{d}}
\label{L_x_eq}
\end{equation}
which is non-convex, with a negative minimum in $d_1=(2a/b)^2$, a positive divergence for $d \to 0$ and a vanishing cost for $d \to \infty$ (see Fig. \ref{graphic_summary}b). 
Here, the argument of the cost function is the agent's mean distance to its nearest neighbours $d$ (we discuss other choices below). We fix the coefficients $a=1$, $b=2$, thus fixing the minimum of the potential in $\bar{d}=1$.
The cost/reward for the state-action pair is then:
\begin{equation}
    c_i^{t+1} = \mathcal{L}(d_i^{t+1}) - \mathcal{L}(d_i^t),
    \label{c_i_t}
\end{equation}
where the cost functions are evaluated before (at time $t$) and after (at time $t+1$) the entire group of $N$ agents updated its position and heading according to Eqs. (\ref{s_t}) and (\ref{r_t}). The difference $c_i^t$ in Eq. \eqref{c_i_t} can be, of course, both negative (reward) or positive (cost). 

First consider the decentralized training (DT) paradigm, where each agent is characterized by a different $Q_i$ matrix. At each time-step $t$, agent $i$ chooses an action $\beta_i$ according to its state $d_i$, evolves its position and heading via Eqs. (\ref{r_t})-(\ref{s_t}), and computes the corresponding cost (\ref{c_i_t}). The $Q_i$-matrix element corresponding to a chosen state-action couple $(d, \beta)$ is updated according to the following update rule 
\begin{equation}
    Q_i^{t+1}(d, \beta) = Q_i^t(d, \beta) + \alpha(d, \beta)[c_i^{t+1} - Q_i^t(d, \beta)],
    \label{Qt}
\end{equation}
where we initialize the $Q_i^0$ matrices to 0 and the learning rate $\alpha_i$ decays as training proceeds according to
\begin{equation}
     \alpha_i(d, \beta) = \frac{\alpha_0}{(1+\Omega_i(d, \beta)/\Omega_0)^{\omega}},
     \label{alpha_sa_i}
\end{equation}
with $\alpha_0\ll 1$ and $\omega \in (0.5,1)$ \cite{singh2000convergence} the so-called learning exponent. The frequency matrix $\Omega_i(d, \beta)$ is the number of times the agent visited that particular state-action pair during the entire learning procedure and $\Omega_0$ is a frequency normalization. 
When an agent selects a state-action couple $(d,\beta)$ the information it will learn will be weighted less the more often that state-action couple has been visited. This ensures the convergence of the $Q$-matrix towards a stationary state.\\

The goal of RL agents is to minimize the cumulative
cost they collect over time, and typically this may be done by 
choosing the current optimal action via Eq. (\ref{optimal}).
On the other hand, agents need to explore all the possible combinations of states and actions in order to find the real optimum. The \textit{exploration-exploitation tradeoff} is a central topic in RL algorithms and, in order to be correctly addressed, one needs to balance the two paradigms correctly. For this reason we decided to induce exploration of new state-action pairs using the so called $\epsilon-greedy$ exploration scheme.
At each time step, agents choose the current optimal action (\ref{optimal}) with probability $p=1-\epsilon_i(d,\beta)$ or, with complementary probability $\epsilon_i(d,\beta)$
 a random action $\beta$ chosen with equal probability between the $n_a$ different options. The {\it exploration rate $\epsilon_i$} is a function of state-actions pairs and decreases during learning according to the decay rule
\begin{equation}
    \epsilon_i(d,\beta) = (1+\Omega_i(d,\beta)/\Omega_0)^{-\nu},
    \label{epsilon_sa}
\end{equation}
with an exploration exponent $\nu \in (0,1)$ \cite{singh2000convergence}. Random exploration avoids the training dynamics to be stuck in sub-optimal minima; analogously to the learning rate, the exploration rates decays towards zero as state action pairs are repeatedly visited, ensuring convergence of the learning algorithm.\\
Learning is typically organized in $n_e$ training episodes of $T$ time-steps (in a time steps all agent evolve once with a synchronous dynamics), with the initial positions and headings of the agents reinitialized at the beginning of each episode. The $Q$-matrices, learning rates and exploration rates, of course, are not reinitialized every episode. At the end of learning, one expects the $Q$-matrices to have practically converged towards a stationary state and one can define a learned dynamics where agents always select the optimal action for each given state via Eq. (\ref{optimal}).\\

Centralizing training (CT) follows the same procedure, but in this case agents share the same information during learning, with single $Q$ and frequency matrices, $Q_i = Q$, $\Omega_i=\Omega$ and learning and exploration rates, $\alpha_i=\alpha$, $\epsilon_i=\epsilon\,\,,\,\forall i$.

\section{Results}
Learning is performed over $n_e$ training episodes, each of a duration of $T=20 N$ time steps. We discretize actions and states with $\Delta \beta=1/(n_a-1)$ and $n_a=11$ actions and $\Delta d=0.2$ with $n_d=20$ different possible states (the 20th state also encodes for mean neighbours distances $d > 20\Delta d=4$). We have verified that our results are robust with respect to reasonable increases in $n_d$ and $n_a$. For what concerns the learning parameters introduced in the previous section, we have performed a preliminary exploration of different values in the convergence windows of RL theory \cite{singh2000convergence} in order to adapt them to our specific problem. While we cannot claim optimality in a rigorous sense, we have settled for $\alpha_0=0.005$, $\Omega_0=N$ and $\nu=\omega=0.7$ for centralized training. These choices seem to grant a sufficiently fast convergence towards a stationary state, with fairly robust results w.r.t. small changes in these parameters. As we will show in Sec. \ref{onset}, decentralized training is more delicate. For sufficiently small flocks, $N<800$, these parameters, with the frequency normalization $\Omega_0=100$ are still efficient, while for larger flocks we obtained better cohesion with $\nu=0.5$ and $\omega=0.99$.

 At the beginning of each training episode the agents' positions and headings are randomly reinitialized (for the positions we uniformly distribute the agents in a circle with a radius $R_0=10 \sqrt{2}$) but of course they keep track of what they learned so far (the Q matrix is not reinitialized). Unless differently stated, we also fix $v_0=0.2$ and $\eta=0.3$ in the agents dynamics. 

\begin{figure}[t]
\includegraphics[width=1.0\linewidth]{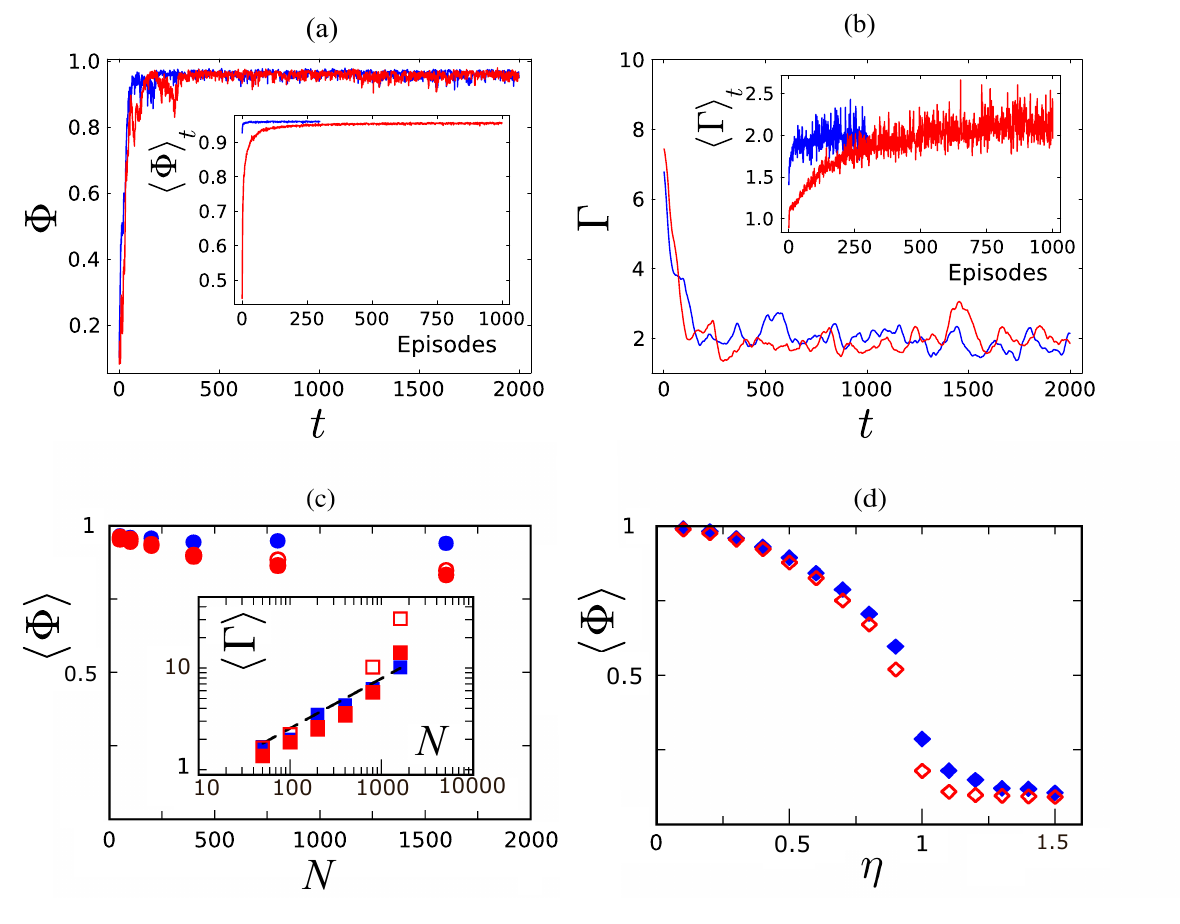}
\caption{ Mean distance to the centre of mass (MDCM) and order parameter (OP) for both centralized (CT) (blue, $n_e=300$) and decentralized (DT) training (red, $n_e=1000$).
  (a) MDCM in the last training episode for a group of $N=100$ agents. (b) OP in the last training episode ($N=100$). Insets in (a) and (b) are averages over single episodes. (c) Finite size scaling of the average OP. Inset: Finite size scaling of the average MDCM, the dashed black line marks the power law $\sim \sqrt{N}$. (d) OP vs. noise amplitude for $N=100$. In panels (c)-(d) open red symbols refer to DT with $\nu=\omega=0.7$, while full red symbols to DT with $\nu=0.5$ and $\omega=0.99$.}
\label{flocking_res}
\end{figure}

\subsection{Onset of cohesive flocking}
\label{onset}
We have verified numerically that our RL algorithm leads to a state of cohesive flocking in a relatively small number of training episodes, both in the CT and DT paradigms, see for instance Video1 in Supplementary. In particular, flocking alignment is quantified as usual via the scalar polar order parameter (OP)
\begin{equation}
    \Phi(t) = \frac{1}{N}\left|\sum_{i=1}^N {\bf s}_i^t\right|\,.
    \label{op}
\end{equation}
For an ordered ensemble of mutually correlated agent headings  $\Phi(t)$ is finite and of order one, while it tends to zero for uncorrelated and disordered headings. In finite flocks, in the latter case one has indeed $\Phi(t) \sim 1/\sqrt{N}$ due to finite  fluctuations. Furthermore, group cohesion can be assessed via the mean distance to the group center of mass (MDCM) which estimates the typical flock size
\begin{equation}
    \Gamma(t) = \frac{1}{N}\sum_{i=1}^N|{\bf r}_i^t-{\bf r}_{cm}^t|
    \label{Gamma}
\end{equation}
with
\begin{equation}
    {\bf r}_{cm}^t = \frac{1}{N}\sum_{i=1}^N {\bf r}_i^t\,.
\end{equation}
Cohesive groups will be therefore characterized by a finite stationary $\Gamma(t)$, which will otherwise grow unbounded in the absence of group cohesion. For instance, for purely alignment interactions with topological neighbours one has $\Gamma(t) \sim \sqrt{t}$ \cite{ginelli2010relevance}.

We first consider groups of $N=100$ agents. The MDCM and OP timeseries during the last episode of training are (respectively) reported in the main panels of Fig. \ref{flocking_res}a-b. Blue curves have been obtained by centralized training, while red ones via decentralized one. Cohesive collective motion, with $\Gamma \approx 2$ and $\langle \Phi \rangle_t \approx 0.95$, is obtained with both training protocols. This high ordered regime is indeed characteristic of starling flocks observed in the wild \cite{cavagna2008new}.
Training is notably efficient, as evidenced by the single-episode averages $\langle \Gamma \rangle_t$ and $\langle \Phi \rangle_t$ of MDCM and OP -- reported in the insets of Fig. \ref{flocking_res}a-b -- 
that settle to their final values after around 30 episodes for CT. DT, on the other hand, requires more training episodes to reach the final trained state, especially for what concerns the saturation of the typical size $\Gamma$ to a stationary value.

We established the scaling of our training algorithm with the number of agents testing groups up to $N=1600$ agents (a typical size for starling flocks observed in the field \cite{cavagna2008new}). 
As shown in Fig. \ref{flocking_res}c, the average OP at the end of training for CT converges towards a finite asymptotic value $\Phi \approx 0.94$, while the flock size $\Gamma$ grows like $\sqrt{N}$, signaling that the flock density remains constant as the group size is varied. Decentralized training, on the other hand, seems to be less robust, with $\Phi \approx 0.84$ and a large flock size $\Gamma$ signaling a partial loss of cohesion at $N=1600$. This is due to the increasing difficulties of individual agents to explore the full range of states $d_i$ in a finite time when $N$ is increased.  In practice, some agents do not learn an optimal policy in a reasonable training time.
This difficulty may be partially mitigated by a change of the exploration ($\nu=0.5$) and learning ($\omega=0.99$) exponents, a choice that increases the late time exploration rate at the expense of a faster decreasing learning rate. As shown in Fig. \ref{flocking_res}, these values lead to nearly identical results for $N<800$ but result in better cohesion at larger system sizes. A more careful analysis of the finite size scaling of training in the DT regime is left for future work.

We have also verified that the order parameter decreases with increasing noise amplitude, with a transition towards swarming behavior ($\Phi \approx 1/\sqrt{N}$) at $\eta_c \approx 1$. A detailed investigation of this transition is however beyond the scope of this work.

\subsection{Flock Structure}

We characterized the structure of cohesive flocks, showing that our flocks behave as a rather structureless nonequilibrium liquid, with a finite 
neighbours exchange rate \cite{mora2016local}
and a pair distribution function qualitatively matching experimental observations in starling flocks \cite{ballerini2008empirical}.

The relative positions of birds in a flock are not fixed, and they are known to exchange neighbours similarly to what happens in an ordinary liquid. Such an exchange can be quantified by the mixing rate \cite{cavagna2014dynamical}
\begin{equation}
\mu = \frac{1}{N}\sum_i^N \frac{1}{m_i} \sum_j^N \left|W_{ij}(t+1)-W_{ij}(t)\right| 
\end{equation}
with the adjacency matrix $W_{ij}(t) = 1$ if agents $i$ and $j$ are topological neighbours at time $t$ and 0 otherwise. A non-zero mixing rate implies neighbour exchange and an irreversible, nonequilibrium dynamics \cite{ginelli2016physics}. Indeed, it can be shown that the entropy production rate of Vicsek-like models is proportional to such a mixing rate \cite{ferretti2022signatures}. The mixing rates measured in our trained flocks are small but finite, $\mu \approx 5 \cdot 10^{-2}$ for the typical parameters considered in this work. Particles do flow, but their neighbour rearrangement occurs on a much slower timescale than local  alignment, consistently with the state of local quasi-equilibrium observed in starling flocks \cite{mora2016local}.

\begin{figure}[b]
\includegraphics[width=1.0\linewidth]{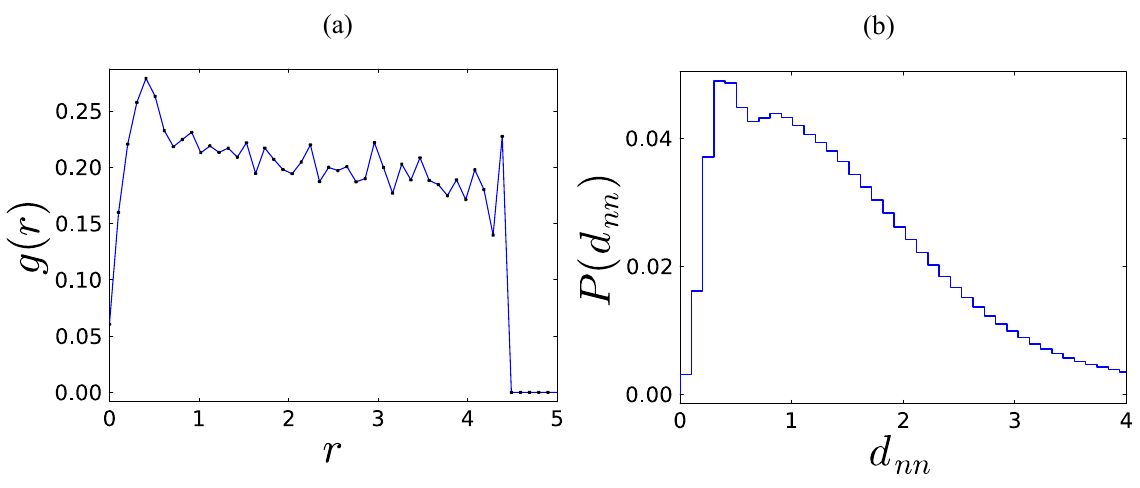}
\caption{(a) Radial pair distribution function measured for a single configuration in a system of $N=400$ agents trained in the CT regime. Here $\Delta r = 0.05$. (b) Probability distribution of the distance of the nearest neighbours obtained from the last episode of the same simulation.} 
\label{gofr}
\end{figure}

The structure of a group can be further investigated by the {\it radial pair distribution function} $g(r)$ which measures the typical probability of finding a particle at distance $r$ from a reference particle. In liquid theory \cite{hansen2013theory} it is typically used to quantitatively characterize the degree of spatial order(gas/liquid/solid) in aggregates of particles.

To compute it, one has to count the number of particles in an annulus of radius $r$ and thickness $\Delta r$ centered on a reference particle and normalize by the annulus area. The result is further averaged over the $n_c(r)$ different reference particles,
\begin{equation}
    g(r) = \frac{1}{2\pi r \Delta r}\frac{1}{n_c(r)}\sum_{i=1}^{n_c(r)}\sum_{j\neq i}\delta_{\Delta r}(r-r_{ij}).
\end{equation}
where $r_{ij}$ is the distance between particles $i$ and $j$ and $\delta_{\Delta r}$ a Dirac delta discretized over a binning $\Delta r$.
In an infinite system one could simply take $n_c(r) = N$, but in a finite system one should be careful to avoid distortions due to the group borders. In order to do so, for any given $r$ one should only consider reference particles whose distance from the border is larger than $r$ \cite{cavagna2008starflag}. Here we define the border particles via an $\alpha$-shape algorithm \cite{edelsbrunner2003shape} with a scale parameter equal to 3 times the mean neighbour distance $d_{nn} = \langle d_i \rangle$.

A typical pair distribution function for a trained flock of $N=400$ agents is reported in Fig.~\ref{gofr}a. It presents a single peak at $r \approx 0.5$ and no further structure, in qualitative agreement with experimental observation in starling flocks \cite{cavagna2008new, camperi2012spatially} whose $g(r)$ also lacks the oscillating behavior of typical liquids. The only peak we observe is a consequence of a well defined peak in the mean neighbour distance distribution (see Fig.~\ref{gofr}b) and is due to the effective repulsion discussed in Sec. \ref{SecQ} and a further repulsive effect due to noisy fluctuations in the orientation \cite{preparation}.

\subsection{Analysis of the Q-matrix} \label{SecQ}
In order to discuss in detail the cohesive flocking rule learned by our agents, we now analyze the trained $Q$-matrix obtained after $n_e$ episodes.

\begin{figure}[!ht]
\includegraphics[width=1.0\linewidth]{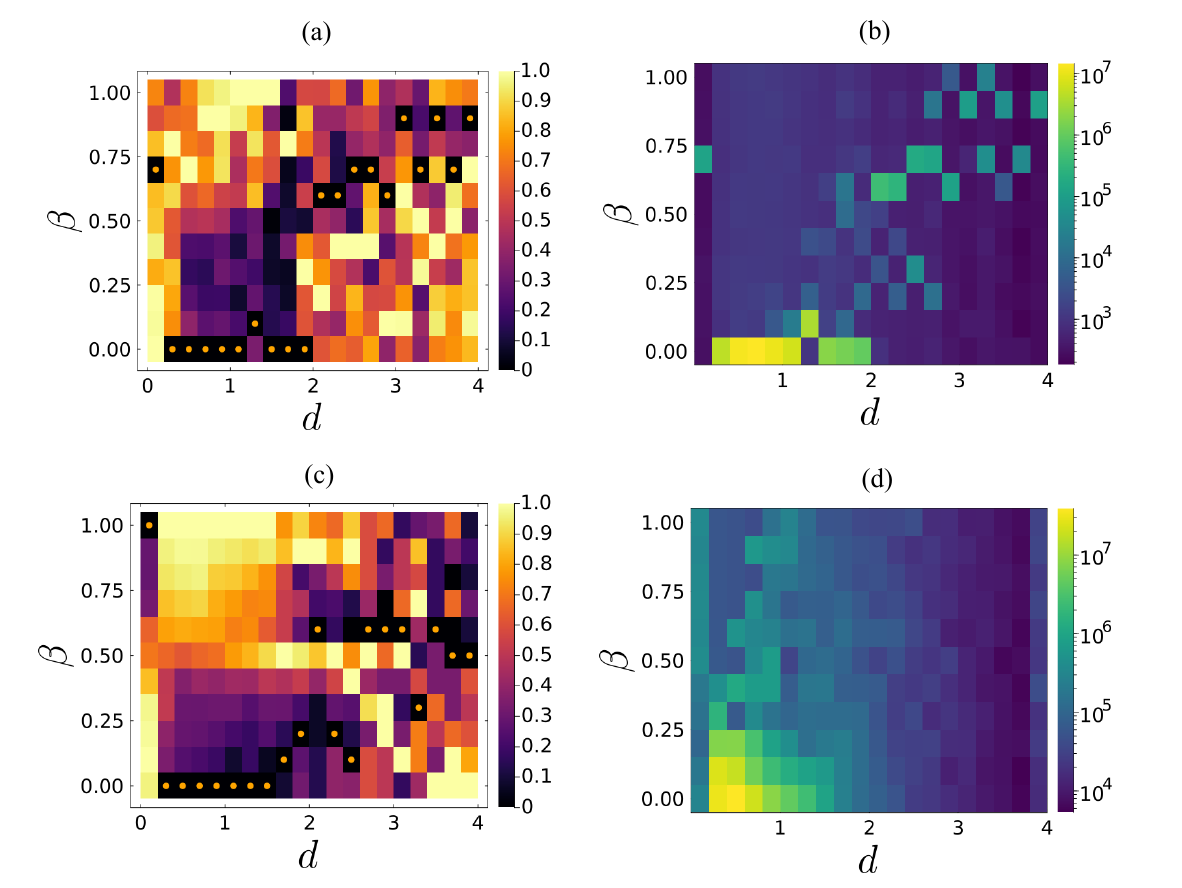}
\caption{\textbf{(a)} Trained $Q$-matrix of the agents for CT after $n_e=300$ episodes. Dots mark the state-action pairs that minimize the matrix for each state. For better visualization, we show the (state-by-state) normalized matrix $\tilde{Q}(d,\beta)\equiv [Q(d,\beta)-Q^m(d)]/[Q^M(d)-Q^m(d)] \in [0,1]$, where $Q^m(d)=\mbox{min}_\beta\{Q(d,\beta)\}$ and $Q^M(d)=\mbox{max}_\beta\{Q(d,\beta)\}$ are respectively the minimum and the maximum values for the state $d$. (b) Final frequency matrix $\Omega$ of the state-action pairs for CT. (c) Trained $Q$-matrix $\langle Q_i \rangle_i$, averaged over all agents for DT after $n_e=1000$ episodes, normalized as in (a). (d) Final total frequency matrix $\sum_i \Omega_i$ for DT. In panels (a) and (c) the optimal policies (see Eq.~\ref{optimal}) are marked by orange dots. $N=100$ in all panels.}
\label{Q_matrix}
\end{figure}

The (normalized) final $Q$-matrices for CT and DT are shown (respectively) in Fig.~\ref{Q_matrix}a and Fig.~\ref{Q_matrix}c for $N=100$. 
They are characterized by two main regimes, a first one for states $d>d^* \approx 2$ and a second one at shorter distances, $d<d^*$.
At large mean distances to neighbours, the optimal actions for the agents, given by Eq.~(\ref{optimal}), involve a combination of alignment with, and attraction to, their neighbours, with $\beta$ seemingly randomly scattered in the interval $(0,1]$. 

Alignment ($\beta\approx 0$), on the other hand, is clearly favoured at smaller distances, $d<d^*$. Contrary to the large distances case, we have verified that strong alignment at short distances is essential for the onset of flocking behavior.
The threshold $d^*$ between these two regimes may slightly change between different training runs due to the stochastic nature of the $\epsilon$-greedy scheme, but it is not too far from the cost function inflection point (where the slope changes sign) located at $d=64a^2/(9 b^2)=16/9$.
It is also interesting to note that the onset of these two different strategies is somehow reminiscent of the distance-dependent behavioral zones assumed in various early flocking models \cite{reynolds1987flocks, couzin2002collective, aoki1980analysis, huth1992simulation}.  

Note finally the exception of the first state, $d<\Delta d=0.2$, where attraction seems to be the optimal action in both CT and DT. 
This rather puzzling result is a consequence of the discrete nature of our dynamics: when $d<\bar{d}$ the cost function rewards actions that increase the mean distance to neighbours $d$ and penalizes those that further reduce it. For $d>v_0$ this simply rules out attraction, resulting in agents with commonly aligned headings that separate slowly due to noisy fluctuations in their individual headings. However, when extremely close, 
$d\leq v_0$, strong attraction may lead to large differences in neighbours' heading, so that nearby agents may easily pass each other and find themselves, after a single time-step, separated by a larger distance. In
practice, at distances smaller than $v_0$, large $\beta$ may act like an effective repulsion
and be favored by the diverging cost function. See also Section III.D for further evidence

We have indeed verified that, for a reduced speed $v_0 \ll \Delta d$ (in particular, we have chosen $v_0=0.05$), this effect is largely mitigated, with a typical optimal policy for the smallest mean neighbour distance state ($d<\Delta d$) dominated by alignment, $\beta \in [0,0.2]$.
In Section \ref{policy} we further show that forcing $\beta=0$ for $d<v_0$ still leads to cohesive flocking.

Note finally that the total CT and DT frequency matrices, reported in Figs.~\ref{Q_matrix}b-d, show that agents predominantly visit state-action pairs characterized by a relatively short distance to neighbours and strong aligning interactions.

\subsection{Policy distillation}
\label{policy}

\begin{figure}[t!]
    \centering
    \includegraphics[width=1\linewidth]{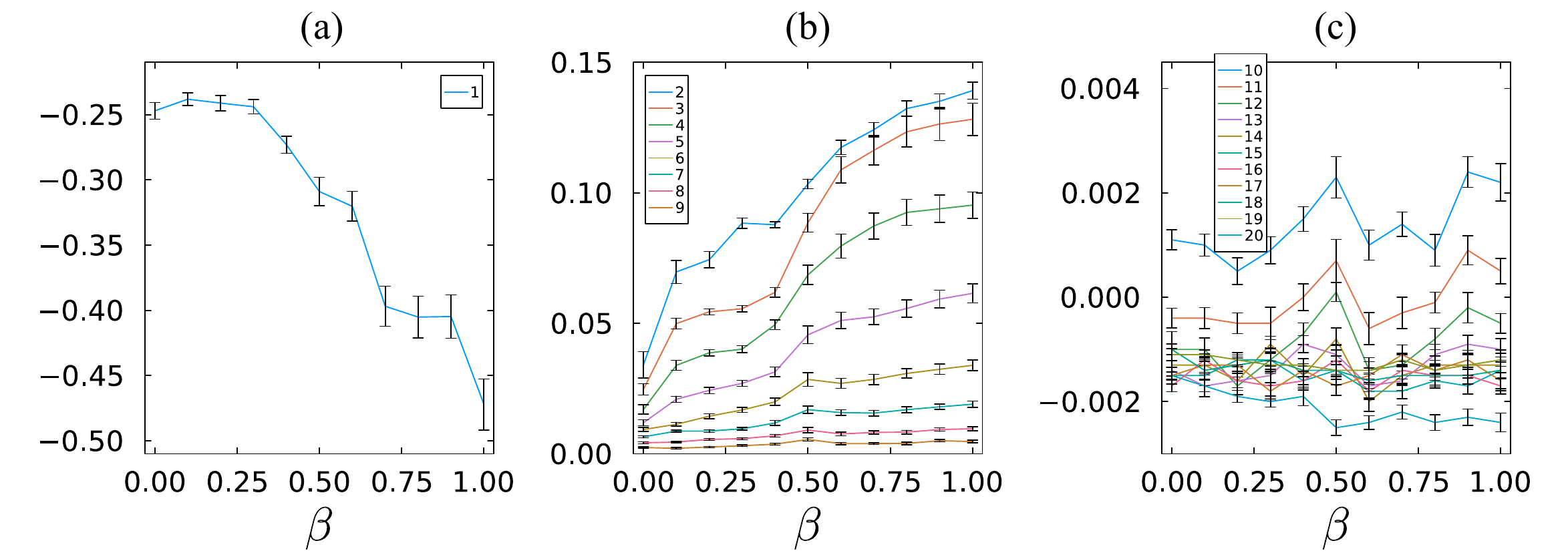}
    \caption{Representation of the average $Q$-matrix $\bar{Q}=\langle Q_i\rangle_i$ trained in the DT scheme for $N=100$ agents. Each curve, labeled by an integer $n$ refers to a specific state $d=n\Delta d$ and shows $\bar{Q}(d, \beta)$ as a function of $\beta$. Error bars represent one standard error. 
    (a) First state, $d<v0$. (b) States with $d<d^*$. (c) States with $d>d^*$. Note that the ratio between the vertical scale of the three panels is 1:2:40.}
    \label{fig:Q_sps_DT}
\end{figure}

A detailed analysis of the trained (and averaged over all agents for the DT scheme) $Q$-matrix state-by-state (see for instance Fig.~\ref{fig:Q_sps_DT}) shows that they are characterized by a well-defined minimum only for small mean neighbour distances. This shows that for $d<d^*$ training strongly converges towards a well-defined  optimal policy of fully aligning interactions (with the only exception of the case $d<v_0$ as discussed in Sec. \ref{SecQ}). 
For $d>d^*$, on the other hand, absolute minima are not well defined, and indeed one can verify that at large distances different training instances do not always converge towards the same optimal policy $\beta^*(d)$. This suggests that, as long as some degree of attractive interaction is present, the exact value of the policy $\beta^*(d)$ is unimportant at large distances. 

\begin{figure}[b!]
 \includegraphics[width=0.99\linewidth]{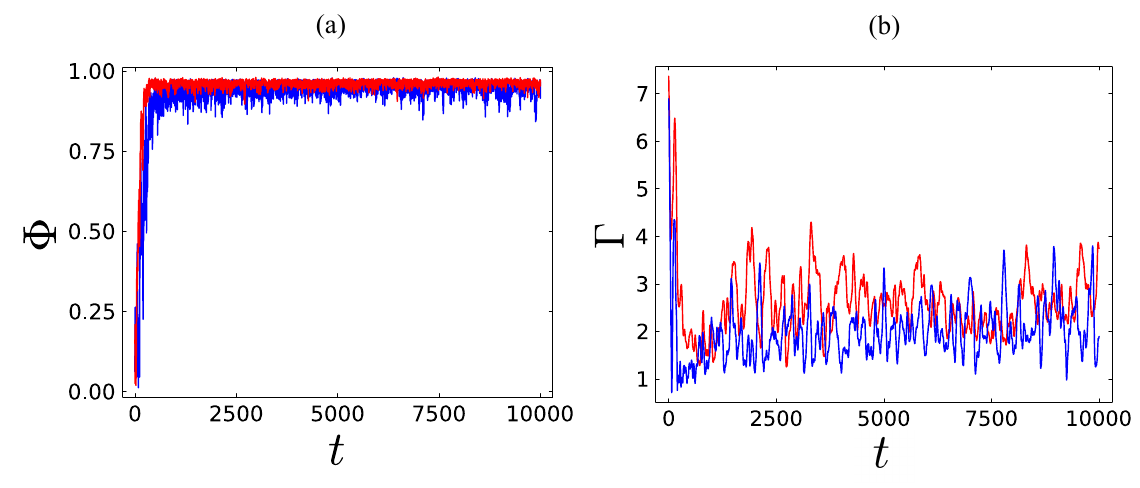}
\caption{Time-series of the order parameter $\Phi$ (a) and flock size $\Gamma$ (b) for $N=100$ agents with two different policy distillations (see text): Random policy for $d>d^*$ (blue curves) and complete aligning interaction for $d<v_0$ (red curves).}
\label{pd}
\end{figure}

To verify this conjecture, we have considered the optimal policy generated by the usual CT training but, for $d>d^*$, we replaced $\beta^*(d)$ with a random value uniformly distributed in $[0,1]$. We have verified that the resulting dynamics shows cohesive flocking and it is qualitatively analogous to the one generated by the original optimal policy, as evidenced by the OP and MDCM time-series shown in red in Fig.~\ref{pd}.

An analogous policy distillation can be performed to remove strong attraction arising for $d<v_0$, as shown in the top left panel of Fig.~\ref{fig:Q_sps_DT}. We have verified that substituting the attractive optimal policy selected by training with pure alignment ($\beta=0$) still results in cohesive flocking (see Fig.~\ref{pd}, red curves), albeit with a slightly larger order parameter and smaller flock size due to the lack of the effective repulsion discussed in Sec. \ref{SecQ}.

\subsection{A swarming counterexample}
We conclude by discussing the role of the cost function (\ref{L_x_eq}) divergence at short distances in the onset of flocking behavior. Such a divergence can be removed by considering instead $\mathcal{L}(d+z)$ with a positive shift $z$. In particular, with the choice $z=\bar{d}=1$ the minimum of the cost function is shifted to $d=0$, completely removing penalties for arbitrarily small agent distances $d$. We have verified, as shown in Fig.~\ref{L_x_fig}, that this choice suppresses flocking and only leads to cohesive swarming (see Video2 in Supplementary). The analysis of the trained $Q$-matrix indeed shows that in this case the orientational dynamics is dominated by attraction ($\beta \approx 1$).

However, a diverging cost function is not a necessary condition for flocking, which also appears for finite but sufficiently large penalties at small $d$. In Fig.~\ref{L_x_fig}b we show how flocking is sustained up to a shift value of around $z_c \approx 0.6$. A detailed investigation of the transition from flocking to swarming as the shift control parameter is varied is beyond the scope of this paper.

\begin{figure}[!t]
\includegraphics[width=1.0\linewidth]{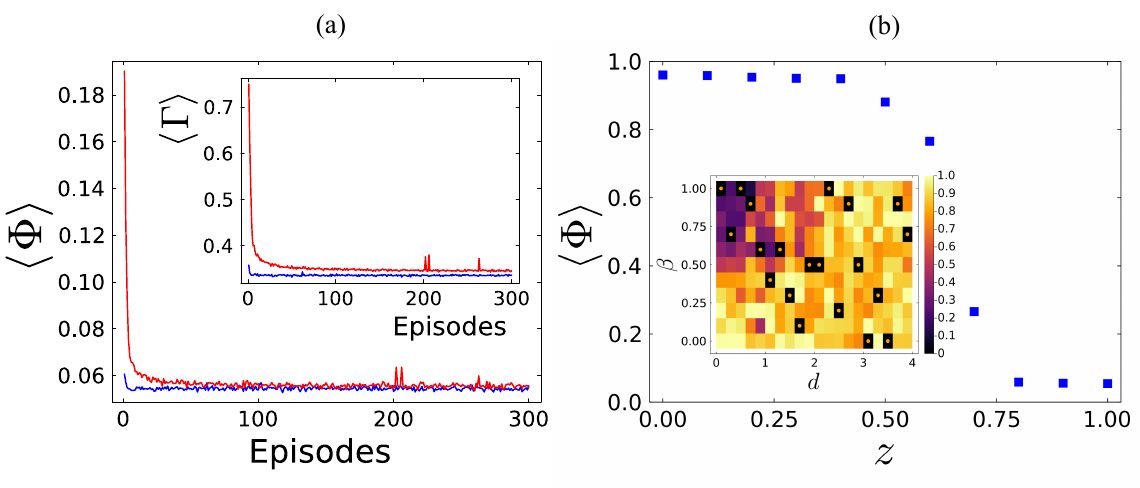}
\caption{(a) Order parameter averaged over single episodes for $z=1$ and CT (blue curves) and DT (red curves) schemes. Inset: Episode averaged MDCM for the same simulations. (b) Stationary order parameter for different shifting parameter $z$ and CT scheme. Inset: trained $Q$-matrix for CT and $z=1$. $N=100$ in both panels.} 
\label{L_x_fig}
\end{figure}

\subsection{Boundary agents and Voronoi based cost}
So far we have evaluated the cost function (\ref{L_x_eq}) on the mean distance to Voronoi neighbors (\ref{c_i_t}). This choice, however, does not distinguish between agents positioned at the boundary and those at the interior of the group. However, one may argue that, from the behavioral standpoint of protection from predators \cite{krause2010important}, agents may have a preference for the latter positions. In order to test this idea, we have chosen the area of the agents' Voronoi cells as the argument of their cost functions. If $\mathcal{A}_i^t$ is the area of the Voronoi cell of agent $i$ at time $t$, the new cost/reward for state-action pairs is now computed as 
\begin{equation}
    c_i^{t+1} = \mathcal{L}(\mathcal{A}_i^{t+1}) - \mathcal{L}(\mathcal{A}_i^t)\,,
    \label{c_i_t2}
\end{equation}
with the convenient choice of cost function parameters $a=\pi/4$ and $b=\sqrt{\pi}$, which fixes the minimum of the cost function at $\bar{\mathcal{A}}=\pi/4$. 

\begin{figure}[!t]
\includegraphics[width=1.0\linewidth]{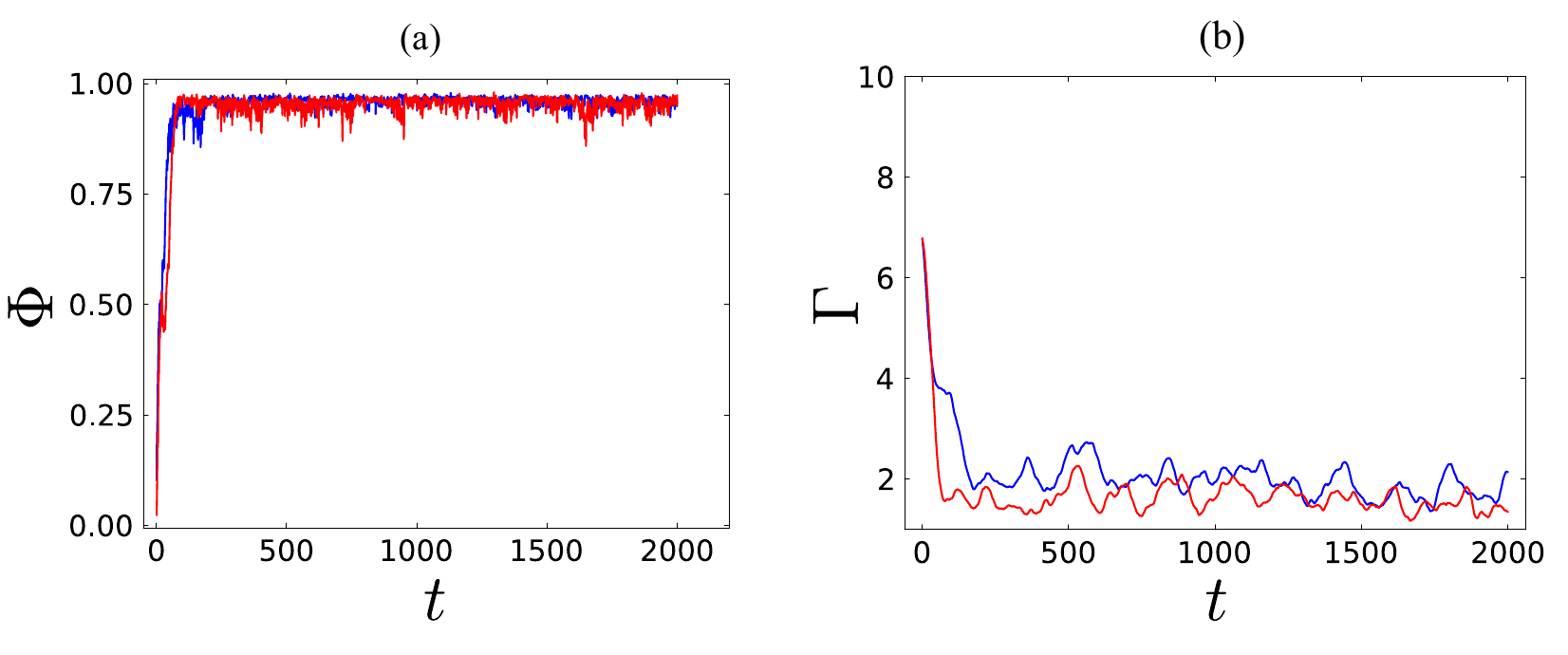}
\caption{Comparison between the optimal policy dynamics of flocks of $N=100$ agents trained in the CT paradigm with cost functions based on mean distance (blue) or Voronoi area (red). (a) Scalar order parameter vs. time. (b) Flock size vs. time.}
\label{Vornoi_vs_d}
\end{figure}

Since Voronoi cells at the group boundary have, by definition, an infinite area, corresponding to a vanishing cost function $\mathcal{L}$, our new choice highlights the cost/reward difference between agents located inside and those at the boundary of the flock. Therefore,  actions resulting in agents moving from the boundary to the interior of the group will be strongly rewarded, while moving from the interior to the border is strongly penalized. Moreover, agents at the boundary get zero cost/reward for any action that does not move them to the inside of the flock, regardless of any change in the distance to their Voronoi neighbours. We have numerically tested this new learning scheme both for centralized and decentralized training, but we have not found any qualitative difference with cost/rewards based on the distance to neighbours, with agents still learning similar optimal policies leading to cohesive flocking, as shown for instance in Fig. \ref{Vornoi_vs_d}.

\section{Discussion}

Using a reinforcement learning framework for a finite group in a boundless space, we demonstrated that cohesive collective motion with high polar order can arise from agents learning to avoid both excessive separation and crowding. While the former requirement is needed to avoid dispersion of the group, the latter condition proved essential for the onset of collective motion: when penalties for small inter-agent distances are removed, the learned strategy shifts toward  attractive interactions and results in a disordered swarming state.

Interestingly, the learned flocking dynamics shows characteristic structural features, such as a slow mixing rate and a structureless pair distribution function, reminiscent of natural bird flocks, reinforcing the biological plausibility of the learned behavior. While our results have been obtained in two spatial dimensions, 
one should remember that, due to the presence of gravity, real birds fly preferentially in the horizontal directions, resulting in a kind of quasi-2d motion \cite{ballerini2008empirical, toner1998flocks}. For this reason, we believe our results may be of direct relevance for real birds.

From a learning perspective, our results are fairly robust under centralized training. The optimal flocking policy adopted by the agents consists of pure alignment at short distances and a combination of alignment and attraction at larger ones. Interestingly, we found that in the latter regime, the precise balance between alignment and attraction is not critical for maintaining cohesive flocking. This implies that multiple policies are equally effective — a redundancy that may contribute to behavioral robustness in biological systems. Decentralized training is almost equally efficient for moderate group sizes, leading to the same optimal policy, but it partially degrades for larger groups. Preliminary analysis suggests that this is due to individual agents not thoroughly exploring the full range of possible states. 

Our findings help clarify the longstanding question of why some species flock while others swarm. Both collective behaviors ensure cohesion in groups of moving agents, but our results suggest that flocking may emerge as the optimal strategy not only for maintaining group cohesion but also for avoiding potentially dangerous collisions.

We finally note that our results may also prove useful beyond animal behavior, particularly in the field of swarm robotics \cite{hamann2018swarm}.
In the future, it may be interesting to generalize these results to  mixed-agent populations and environments with external threats/perturbations in order to extend the result of \cite{cavagna2013boundary} to the full non-equilibrium dynamics learned by our agents.

\section{Acknowledgments}
We thank H. Chat\'e for carefully reading the manuscript and providing helpful feedback. M.B., A.C and F.G. acknowledge support from grant PRIN 2020PFCXPE from MIUR.

\bibliography{biblio}

@article{rackham1942pliny,
  title={Pliny Natural History (The Loeb Classical Library)},
  author={Rackham, H and Jones, WHS and Eichholz, DE},
  journal={London, Cambridge},
  year={1942}
}

@article{parrish1999complexity,
  title={Complexity, pattern, and evolutionary trade-offs in animal aggregation},
  author={Parrish, Julia K and Edelstein-Keshet, Leah},
  journal={Science},
  volume={284},
  number={5411},
  pages={99--101},
  year={1999},
  publisher={American Association for the Advancement of Science}
}

@article{cavagna2010scale,
  title={Scale-free correlations in starling flocks},
  author={Cavagna, Andrea and Cimarelli, Alessio and Giardina, Irene and Parisi, Giorgio and Santagati, Raffaele and Stefanini, Fabio and Viale, Massimiliano},
  journal={Proceedings of the National Academy of Sciences},
  volume={107},
  number={26},
  pages={11865--11870},
  year={2010},
  publisher={National Academy of Sciences}
}

@article{ballerini2008,
  title={Interaction ruling animal collective behavior depends on topological rather than metric distance: Evidence from a field study},
  author={Ballerini, Michele and Cabibbo Nicola and Candelier, Raphael and Cavagna, Andrea and Cisbani, Evaristo and Giardina, Irene and Lecomte, Vivien and Orlandi, Alberto and Parisi, Giorgio and Procaccini, Andrea  and Viale, Massimiliano and Zdravkovic, Vladimir},
  journal={Proceedings of the National Academy of Sciences},
  volume={105},
  number={4},
  pages={11865--11870},
  year={1232-1237},
  publisher={National Academy of Sciences}
}

@inproceedings{reynolds1987flocks,
  title={Flocks, herds and schools: A distributed behavioral model},
  author={Reynolds, Craig W},
  booktitle={Proceedings of the 14th annual conference on Computer graphics and interactive techniques},
  pages={25--34},
  year={1987}
}

@article{vicsek1995novel,
  title={Novel type of phase transition in a system of self-driven particles},
  author={Vicsek, Tam{\'a}s and Czir{\'o}k, Andr{\'a}s and Ben-Jacob, Eshel and Cohen, Inon and Shochet, Ofer},
  journal={Physical review letters},
  volume={75},
  number={6},
  pages={1226},
  year={1995},
  publisher={APS}
}

@article{ginelli2016physics,
  title={The physics of the Vicsek model},
  author={Ginelli, Francesco},
  journal={The European Physical Journal Special Topics},
  volume={225},
  number={11},
  pages={2099--2117},
  year={2016},
  publisher={Springer}
}

@article{ginelli2010relevance,
  title={Relevance of metric-free interactions in flocking phenomena},
  author={Ginelli, Francesco and Chat{\'e}, Hugues},
  journal={Physical review letters},
  volume={105},
  number={16},
  pages={168103},
  year={2010},
  publisher={APS}
}

@article{gregoire2004onset,
  title={Onset of collective and cohesive motion},
  author={Gr{\'e}goire, Guillaume and Chat{\'e}, Hugues},
  journal={Physical review letters},
  volume={92},
  number={2},
  pages={025702},
  year={2004},
  publisher={APS}
}

@article{chate2008modeling,
  title={Modeling collective motion: variations on the Vicsek model},
  author={Chat{\'e}, Hugues and Ginelli, Francesco and Gr{\'e}goire, Guillaume and Peruani, Fernando and Raynaud, Franck},
  journal={The European Physical Journal B},
  volume={64},
  number={3},
  pages={451--456},
  year={2008},
  publisher={Springer}
}

@article{couzin2002collective,
  title={Collective memory and spatial sorting in animal groups},
  author={Couzin, Iain D and Krause, Jens and James, Richard and Ruxton, Graeme D and Franks, Nigel R},
  journal={Journal of theoretical biology},
  volume={218},
  number={1},
  pages={1--11},
  year={2002},
  publisher={Elsevier}
}

@article{pearce2014role,
  title={Role of projection in the control of bird flocks},
  author={Pearce, Daniel JG and Miller, Adam M and Rowlands, George and Turner, Matthew S},
  journal={Proceedings of the National Academy of Sciences},
  volume={111},
  number={29},
  pages={10422--10426},
  year={2014},
  publisher={National Academy of Sciences}
}

@article{giraldo2025active,
  title={Active Matter Flocking via Predictive Alignment},
  author={Giraldo-Barreto, Julian and Holubec, Viktor},
  journal={arXiv preprint arXiv:2504.07778},
  year={2025}
}

@article{barto2021reinforcement,
  title={Reinforcement learning: An introduction. by richard’s sutton},
  author={Barto, Andrew G},
  journal={SIAM Rev},
  volume={6},
  number={2},
  pages={423},
  year={2021},
  publisher={SIAM}
}

@article{durve2020learning,
  title={Learning to flock through reinforcement},
  author={Durve, Mihir and Peruani, Fernando and Celani, Antonio},
  journal={Physical Review E},
  volume={102},
  number={1},
  pages={012601},
  year={2020},
  publisher={APS}
}

@article{borra2021optimal,
  title={Optimal collision avoidance in swarms of active Brownian particles},
  author={Borra, Francesco and Cencini, Massimo and Celani, Antonio},
  journal={Journal of Statistical Mechanics: Theory and Experiment},
  volume={2021},
  number={8},
  pages={083401},
  year={2021},
  publisher={IOP Publishing}
}

@article{cavagna2008new,
  title={New statistical tools for analyzing the structure of animal groups},
  author={Cavagna, Andrea and Cimarelli, Alessio and Giardina, Irene and Orlandi, Alberto and Parisi, Giorgio and Procaccini, Andrea and Santagati, Raffaele and Stefanini, Fabio},
  journal={Mathematical biosciences},
  volume={214},
  number={1-2},
  pages={32--37},
  year={2008},
  publisher={Elsevier}
}

@article{wang2023modeling,
  title={Modeling collective motion for fish schooling via multi-agent reinforcement learning},
  author={Wang, Xin and Liu, Shuo and Yu, Yifan and Yue, Shengzhi and Liu, Ying and Zhang, Fumin and Lin, Yuanshan},
  journal={Ecological Modelling},
  volume={477},
  pages={110259},
  year={2023},
  publisher={Elsevier}
}

@article{loffler2023collective,
  title={Collective foraging of active particles trained by reinforcement learning},
  author={L{\"o}ffler, Robert C and Panizon, Emanuele and Bechinger, Clemens},
  journal={Scientific Reports},
  volume={13},
  number={1},
  pages={17055},
  year={2023},
  publisher={Nature Publishing Group UK London}
}

@article{cai2025reinforcement,
  title={Reinforcement Learning for Active Matter},
  author={Cai, Wenjie and Wang, Gongyi and Zhang, Yu and Qu, Xiang and Huang, Zihan},
  journal={arXiv preprint arXiv:2503.23308},
  year={2025}
}

@article{monter2023dynamics,
  title={Dynamics and risk sharing in groups of selfish individuals},
  author={Monter, Samuel and Heuthe, Veit-Lorenz and Panizon, Emanuele and Bechinger, Clemens},
  journal={Journal of Theoretical Biology},
  volume={562},
  pages={111433},
  year={2023},
  publisher={Elsevier}
}

@article{singh2000convergence,
  title={Convergence results for single-step on-policy reinforcement-learning algorithms},
  author={Singh, Satinder and Jaakkola, Tommi and Littman, Michael L and Szepesv{\'a}ri, Csaba},
  journal={Machine learning},
  volume={38},
  number={3},
  pages={287--308},
  year={2000},
  publisher={Springer}
}

@article{mora2016local,
  title={Local equilibrium in bird flocks},
  author={Mora, Thierry and Walczak, Aleksandra M and Del Castello, Lorenzo and Ginelli, Francesco and Melillo, Stefania and Parisi, Leonardo and Viale, Massimiliano and Cavagna, Andrea and Giardina, Irene},
  journal={Nature physics},
  volume={12},
  number={12},
  pages={1153--1157},
  year={2016},
  publisher={Nature Publishing Group UK London}
}

@article{camperi2012spatially,
  title={Spatially balanced topological interaction grants optimal cohesion in flocking models},
  author={Camperi, Marcelo and Cavagna, Andrea and Giardina, Irene and Parisi, Giorgio and Silvestri, Edmondo},
  journal={Interface focus},
  volume={2},
  number={6},
  pages={715--725},
  year={2012},
  publisher={The Royal Society}
}

@article{aoki1980analysis,
  title={An analysis of the schooling behavior of fish: internal organization and communication process},
  author={Aoki, Ichir{\=o}},
  journal={Bull. Ocean. Res. Inst. Univ. Tokyo},
  volume={12},
  pages={1--65},
  year={1980}
}

@article{huth1992simulation,
  title={The simulation of the movement of fish schools},
  author={Huth, Andreas and Wissel, Christian},
  journal={Journal of theoretical biology},
  volume={156},
  number={3},
  pages={365--385},
  year={1992},
  publisher={Elsevier}
}

@article{krause2010important,
  title={Important topics in group living},
  author={Krause, Jens and Ruxton, Graeme},
  journal={Social behaviour: genes, ecology and evolution},
  pages={203--225},
  year={2010},
  publisher={Cambridge University Press Cambridge, UK}
}

@article{ballerini2008empirical,
  title={Empirical investigation of starling flocks: a benchmark study in collective animal behaviour},
  author={Ballerini, Michele and Cabibbo, Nicola and Candelier, Raphael and Cavagna, Andrea and Cisbani, Evaristo and Giardina, Irene and Orlandi, Alberto and Parisi, Giorgio and Procaccini, Andrea and Viale, Massimiliano and others},
  journal={Animal behaviour},
  volume={76},
  number={1},
  pages={201--215},
  year={2008},
  publisher={Elsevier}
}

@article{toner1998flocks,
  title={Flocks, herds, and schools: A quantitative theory of flocking},
  author={Toner, John and Tu, Yuhai},
  journal={Physical review E},
  volume={58},
  number={4},
  pages={4828},
  year={1998},
  publisher={APS}
}

@book{hamann2018swarm,
  title={Swarm robotics: A formal approach},
  author={Hamann, Heiko},
  volume={221},
  year={2018},
  publisher={Springer}
}

@article{cavagna2014dynamical,
  title={Dynamical maximum entropy approach to flocking},
  author={Cavagna, Andrea and Giardina, Irene and Ginelli, Francesco and Mora, Thierry and Piovani, Duccio and Tavarone, Raffaele and Walczak, Aleksandra M},
  journal={Physical Review E},
  volume={89},
  number={4},
  pages={042707},
  year={2014},
  publisher={APS}
}

@article{ferretti2022signatures,
  title={Signatures of irreversibility in microscopic models of flocking},
  author={Ferretti, Federica and Grosse-Holz, Simon and Holmes, Caroline and Shivers, Jordan L and Giardina, Irene and Mora, Thierry and Walczak, Aleksandra M},
  journal={Physical Review E},
  volume={106},
  number={3},
  pages={034608},
  year={2022},
  publisher={APS}
}

@book{hansen2013theory,
  title={Theory of simple liquids: with applications to soft matter},
  author={Hansen, Jean-Pierre and McDonald, Ian Ranald},
  year={2013},
  publisher={Academic press}
}

@article{cavagna2008starflag,
  title={The STARFLAG handbook on collective animal behaviour: Part II, three-dimensional analysis},
  author={Cavagna, Andrea and Giardina, Irene and Orlandi, Alberto and Parisi, Giorgio and Procaccini, Andrea},
  journal={arXiv preprint arXiv:0802.1674},
  year={2008}
}

@article{edelsbrunner2003shape,
  title={On the shape of a set of points in the plane},
  author={Edelsbrunner, Herbert and Kirkpatrick, David and Seidel, Raimund},
  journal={IEEE Transactions on information theory},
  volume={29},
  number={4},
  pages={551--559},
  year={2003},
  publisher={IEEE}
}

@article{preparation,
  author={Brambati, Martino and Baj, Giovanni and Ginelli, Francesco},
  title = {},
  note   = {In preparation.},
  year={2026},
  journal = {}
}

@article{cavagna2013boundary,
  title={Boundary information inflow enhances correlation in flocking},
  author={Cavagna, Andrea and Giardina, Irene and Ginelli, Francesco},
  journal={Physical review letters},
  volume={110},
  number={16},
  pages={168107},
  year={2013},
  publisher={APS}
}

@article{gautrais2012deciphering,
  title={Deciphering interactions in moving animal groups},
  author={Gautrais, Jacques and Ginelli, Francesco and Fournier, Richard and Blanco, St{\'e}phane and Soria, Marc and Chat{\'e}, Hugues and Theraulaz, Guy},
  year={2012},
  publisher={Public Library of Science San Francisco, USA},
  journal = {}
}

\end{document}